\documentclass[aps,pra,twocolumn,groupedaddress,longbibliography]{revtex4}

\usepackage{amsmath}
\usepackage{amssymb}
\usepackage{wrapfig}
\usepackage{mathrsfs}
\usepackage{cases}
\usepackage{braket}
\usepackage{verbatim}
\usepackage{latexsym}
\usepackage{multirow}
\usepackage{color}
\usepackage{graphicx}
\usepackage{url}
\usepackage[toc, page]{appendix}
\usepackage[T1]{fontenc}
\usepackage[
colorlinks=true,
linkcolor=blue,
citecolor=blue,
urlcolor=black,
]{hyperref}

\def\v#1{\mbox{\boldmath $#1$}}

\def\pd#1#2{\frac{\partial #1}{\partial #2}}

\newcommand{\Sig}{\sigma}

\begin{document}

\title{Quantum annealing with a nonvanishing final value of the transverse field}

\author{Kohji Nishimura}
\affiliation{Department of Physics, Tokyo Institute of Technology, Oh-okayama, Meguro-ku, Tokyo 152-8551, Japan}

\author{Hidetoshi Nishimori}
\affiliation{Department of Physics, Tokyo Institute of Technology, Oh-okayama, Meguro-ku, Tokyo 152-8551, Japan}

\date{\today}

\begin{abstract}
	We study the problem to infer the original ground state of a spin-glass Hamiltonian out of the information from the Hamiltonian with interactions deviated from the original ones. Our motivation comes from quantum annealing on a real device in which the values of interactions are degraded by noise. We show numerically for quasi-one-dimensional systems that the Hamming distance between the original ground state and the inferred spin state is minimized when we stop the process of quantum annealing before the amplitude of the transverse field reaches zero in contrast to the conventional prescription.  This result means that finite quantum fluctuations compensate for the effects of noise, at least to some extent. Analytical calculations using the infinite-range mean-field model support our conclusion qualitatively.
\end{abstract}


\maketitle

\section{Introduction}

Quantum annealing (QA) is a quantum-mechanical metaheuristic for solving combinatorial optimization problem \cite{QA,Brooke1999,Farhi2001,Santoro2002,Santoro2006,Das2008,Morita2008,Tanaka_book2017}. One controls the strength of quantum fluctuations through the amplitude of the transverse field applied to the Ising model with complex interactions.  Since combinatorial optimization problems can generally be mapped to the Ising model with complex interactions \cite{lucas2014}, i.e. Ising spin glasses, it is important to investigate the performance of QA as a potentially powerful method to find solutions of combinatorial optimization problems. Interest in QA has been accelerated by the hardware implementation of QA, the D-Wave machine \cite{Johnson2011}. The device is, nevertheless, still far from perfect, and one of the problems is the control error, imperfections in the setting of parameter values at the hardware level. Accordingly, the device attempts to find the ground state of the wrong Hamiltonian even if the process of QA is performed perfectly. Hence it is important to devise strategies to mitigate this and other difficulties. One of the possible approaches is to apply quantum error-correcting codes \cite{jordan2006error,PhysRevLett.100.160506,PhysRevA.86.042333,Young:13,Sarovar:2013kx,Ganti:13,Young:2013fk,Bookatz:2014uq,Marvian:2014nr,Jiang:2015kx,Marvian:2016kb,Marvian-Lidar:16,PAL:13,PAL:14,Vinci:2015jt,Mishra:2015,vinci2015nested,Matsuura2016,Matsuura2017}, but this method consumes extra qubits to maintain robustness against noise.

In a recent paper \cite{Nishimura2016}, we showed that applying proper thermal fluctuations to the Ising model with noisy interactions has a potential to infer the original ground state more faithfully than the naive QA at zero temperature. Although this strategy can be applied to a quantum annealer in principle, there remain some drawbacks from a practical perspective. One of the serious problems is that one cannot control the temperature of the device as a parameter to optimize the performance of QA. In contrast, one can control the strength of the transverse field, and thus the field strength may be used as a tunable parameter to extract the maximum possible information from the output of the system.

In the present paper, we  therefore make use of the strength of the transverse field to infer the original ground state of the Ising spin glass with noise in interactions. We show that the best result is achieved when we keep the strength of the transverse field finite at the end of the annealing process in marked contrast to the conventional prescription to reduce the field to zero.  This result may be viewed as a generalization of the conclusion given in \cite{Otsubo2012} for the mean-field model with non-random ferromagnetic interactions in the original Hamiltonian to the cases of low-dimensional as well as the mean-field models with random original interactions.

This paper is organized as follows. We formulate the problem in Sec. \ref{sec:formulation}. Results of numerical calculations are given in Sec. \ref{sec:numericalresult}. Section \ref{sec:mfa} explains an analytical approach by the mean-field model, and is followed by conclusion in Sec. \ref{sec:conclusion}.  Some of the details are delegated to the Appendix.

\section{Formulation of the problem} 
\label{sec:formulation}
Let us define the problem to be discussed in this paper. The goal is to infer the original ground state of a Hamiltonian with interactions $J_{i_1, \ldots, i_p}$, which follow the Gaussian distribution, by using another Hamiltonian whose interactions have been degraded by noise from the original $J_{i_1, \ldots, i_p}$ to $\tilde{J}_{i_1, \ldots, i_p}$ as specified later.

The original spin-glass Hamiltonian is written with general many-body interactions as
\begin{align}
	\label{eq:origH}
	H(\v{\hat{\Sig}}) = -\sum_{i_1, \ldots, i_p}J_{i_1, \ldots, i_p}\hat{\Sig}^z_{i_1}\cdots \hat{\Sig}^z_{i_p}
\end{align}
where $\v{\hat{\Sig}}$ is a set of Pauli operators $\{\hat{\Sig}_i^x,\hat{\Sig}_i^y,\hat{\Sig}_i^z\}_{i=1}^N$ and $p\ (\geq 2)$ is an integer. The original interactions $J_{i_1, \ldots, i_p}$ are generated from the Gaussian distribution with mean unity and variance $\sigma^2$ \footnote{The symbol $\sigma^2$ for variance should not be confused with the Pauli operator, the latter having a hat, such as $\hat{\Sig}_i^z$}. The spin configuration at site $i$ can be extracted from the ground state $\ket{\psi_0}$ of Eq. (\ref{eq:origH}) as
\begin{align}
	\mathcal{S}^{(0)}_i := \braket{\psi_0|\hat{\Sig}^z_i|\psi_0} (=\pm 1).
\end{align}
We assume that there is no degeneracy in the ground state, which is very plausible because of the continuous nature of the Gaussian distribution.

We next introduce the degraded interaction $\tilde{J}_{i_1, \ldots, i_p}$ by adding a noise term $\xi_{i_1, \ldots, i_p}$ to the original interaction,
\begin{align}
	\tilde{J}_{i_1, \ldots, i_p} = J_{i_1, \ldots, i_p} + \xi_{i_1, \ldots, i_p}.
\end{align}
We assume that the noise $\xi_{i_1, \ldots, i_p}$ follows the Gaussian distribution with zero mean and variance $\gamma^2$. Remember that the problem is to identify the spin configuration closest to the ground state of the original Hamiltonian Eq. (\ref{eq:origH}) from the noisy Hamiltonian with interactions $\{\tilde{J}_{i_1, \ldots, i_p}\}$ by adjusting the transverse field  as defined later in Eq. (\ref{eq:noisyH}). The inferred spin configuration is
\begin{align}
	\label{eq:SGamma}
	\mathcal{S}^{(\Gamma)}_i = \mathrm{sgn} \braket{\tilde{\psi}_\Gamma|\hat{\Sig}^z_i|\tilde{\psi}_\Gamma},
\end{align}
where $\ket{\tilde{\psi}_\Gamma}$ stands for the ground state of the noisy Hamiltonian $\tilde{H}(\v{\hat{\Sig}},\Gamma )$,
\begin{align}
	\label{eq:noisyH}
	\tilde{H}(\v{\hat{\Sig}}, \Gamma) = -\sum_{i_1, \ldots, i_p}\tilde{J}_{i_1, \ldots, i_p}\hat{\Sig}^z_{i_1}\cdots \hat{\Sig}^z_{i_p}-\Gamma \sum_{i=1}^N\hat{\Sig}^x_{i}.
\end{align}
Notice that we are interested only in the sign of the spin configuration, not the magnitude, as shown in Eq. (\ref{eq:SGamma}).
As a measure of the distance between these two spin configurations $\mathcal{S}^{(0)}_i$ and $\mathcal{S}^{(\Gamma)}_i$, we introduce the overlap as a function of the magnitude of the transverse field $M(\Gamma)$,
\begin{align}
	\label{eq:MGamma}
	M(\Gamma) = \int \prod dJ P(J)\prod d\xi P(\xi) \mathcal{S}^{(0)}_i \mathcal{S}^{(\Gamma)}_i,
\end{align}
where $P(J)$ and $P(\xi)$ denote the probability measures of the Gaussian distribution, and the products run over $\{J_{i_1, \ldots, i_p}\}$ and $\{\xi_{i_1, \ldots, i_p}\}$, respectively. In the limit of a large system size $N\to\infty$, the average over the interactions in Eq. (\ref{eq:MGamma}) is equivalent to the arithmetic average over all indices,
\begin{equation}
    M(\Gamma) = \frac{1}{N}\sum_{i=1}^N\mathcal{S}^{(0)}_i \mathcal{S}^{(\Gamma)}_i,
\end{equation}
due to the self-averaging property \cite{nsmrbook}. The overlap is closely related to the Hamming distance.  When the overlap goes to unity, the corresponding Hamming distance goes to zero, and we retrieve the original ground state completely, $\mathcal{S}^{(0)}_i  = \mathcal{S}^{(\Gamma)}_i$ at almost all $i$. The optimal inference is accomplished when the overlap $M(\Gamma)$ becomes maximum. We are interested in when this condition is satisfied.

When the variance of the original distribution is zero, $\sigma = 0$, the original Hamiltonian of Eq. (\ref{eq:origH}) is the simple ferromagnetic Ising model.  The ground state trivially has $\mathcal{S}^{(0)}_i=1$  for all $i$ (and  $\mathcal{S}^{(0)}_i=-1 ~\forall i$ when $p$ is even).  This is the situation discussed in the context of classical error-correcting codes, where all original bits are unity \cite{nsmrbook}. The behavior of the overlap in this case was analyzed in Ref. \cite{Otsubo2012} for a mean-field model, and our contribution is to drop the condition of the uniform ferromagnetic interactions in the original model and treat the general spin glass problem. Also, we numerically study the quasi-one-dimensional system, in addition to the mean-field model.

\section{Numerical analysis} 
\label{sec:numericalresult}
\begin{figure}
	\centering
	\includegraphics[width=\columnwidth]{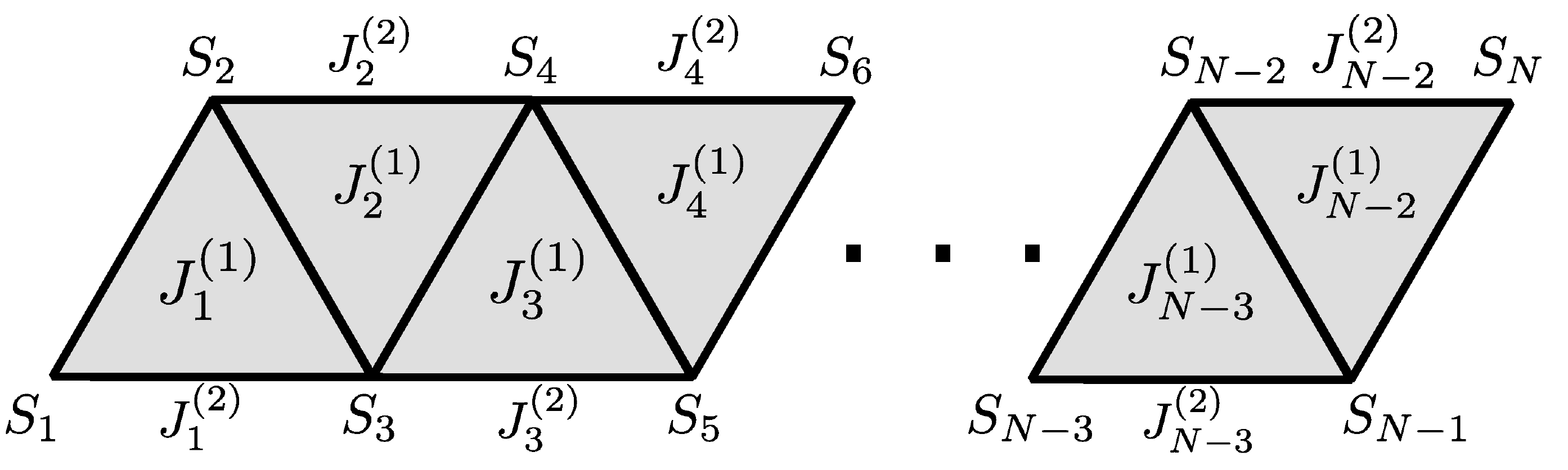}
	\caption{Ising spin glass on a triangular ladder. Each Pauli operator $\hat{\Sig}_i^{z}$ is indicated as $S_i$ in this figure.}
	\label{fig:ladder}
\end{figure}

To calculate the overlap, we follow Refs. \cite{rujan1993,Nishimura2016} to adopt the quasi-one-dimensional triangular ladder with two- and three-body interactions as depicted in Fig.\ref{fig:ladder}. 
The Hamiltonian is
\begin{align}
	\label{eq:ladder}
	H(\v{\hat{\Sig}}) = -\sum_{i=1}^{N-2}\Big(J_i^{(1)}\hat{\Sig}_{i}^{z}\hat{\Sig}_{i+1}^{z}\hat{\Sig}_{i+2}^{z}+J_i^{(2)}\hat{\Sig}_{i}^{z}\hat{\Sig}_{i+2}^{z}\Big).
\end{align}
The ground state of Eq. (\ref{eq:noisyH}) is calculated by the Density Matrix Renormalization Group (DMRG) algorithm \cite{White1992,White1993}. A special care should be taken when performing DMRG in the present case with randomness since this algorithm tends to be stuck in a local minimum for small $\Gamma$ due to the randomness of interactions. To mitigate this problem, we apply a variant of DMRG based on the idea of quantum wavefunction annealing \cite{Rodriguez-Laguna2007, Rodriguez-Laguna2014}, where the transverse field is gradually decreased from a large value to zero in the course of the calculation. The original ground state $\mathcal{S}^{(0)}_i$ is determined by  the Viterbi algorithm \cite{viterbi1967}, which is essentially the numerical transfer matrix method at zero temperature.

The number of spins $N$ is set to 298 and we take the configurational average over the distributions of original interactions and noise by sampling 1200 disorder realizations.

\begin{figure*}
	\centering

	\includegraphics[width=\columnwidth]{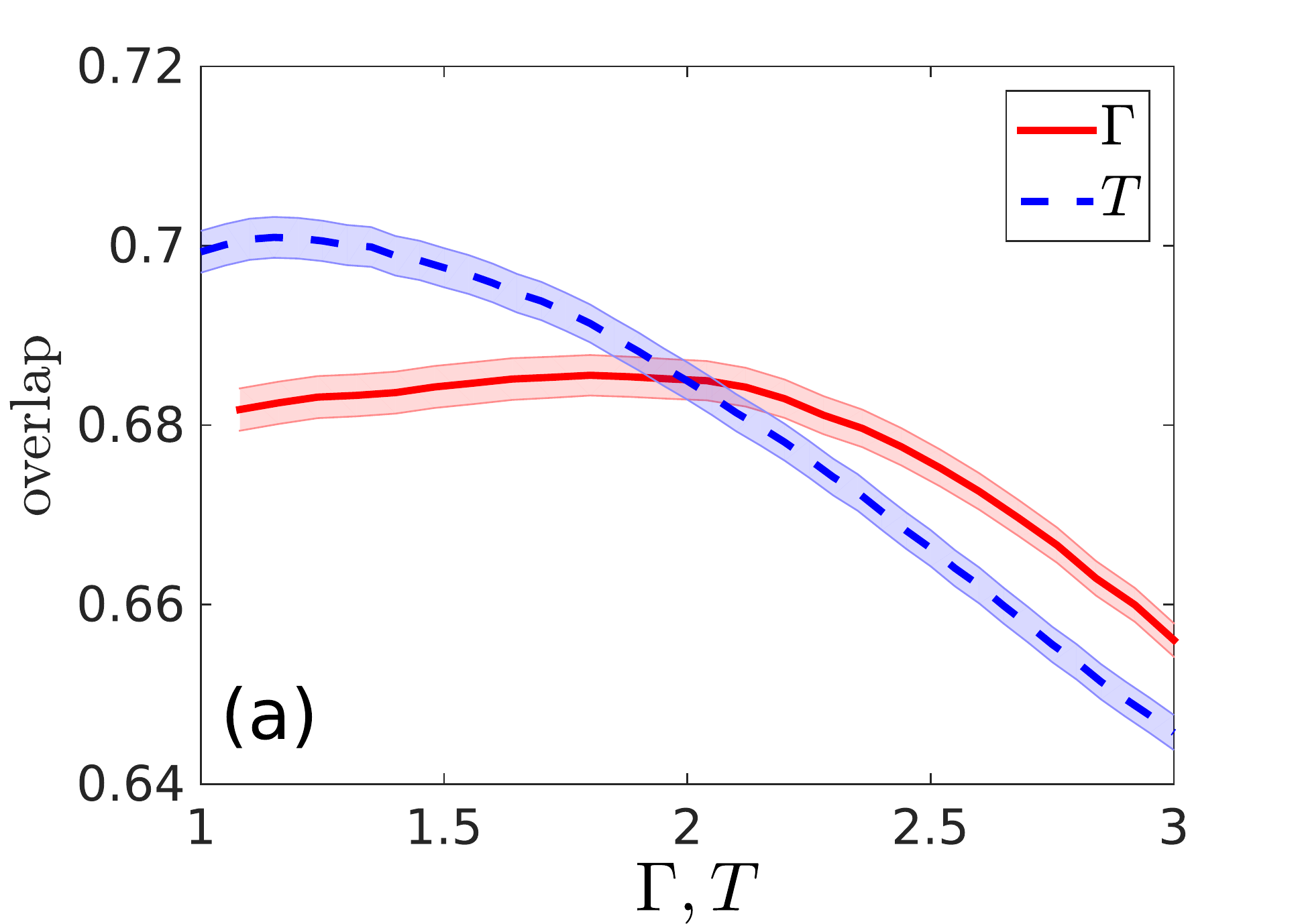}
	\includegraphics[width=\columnwidth]{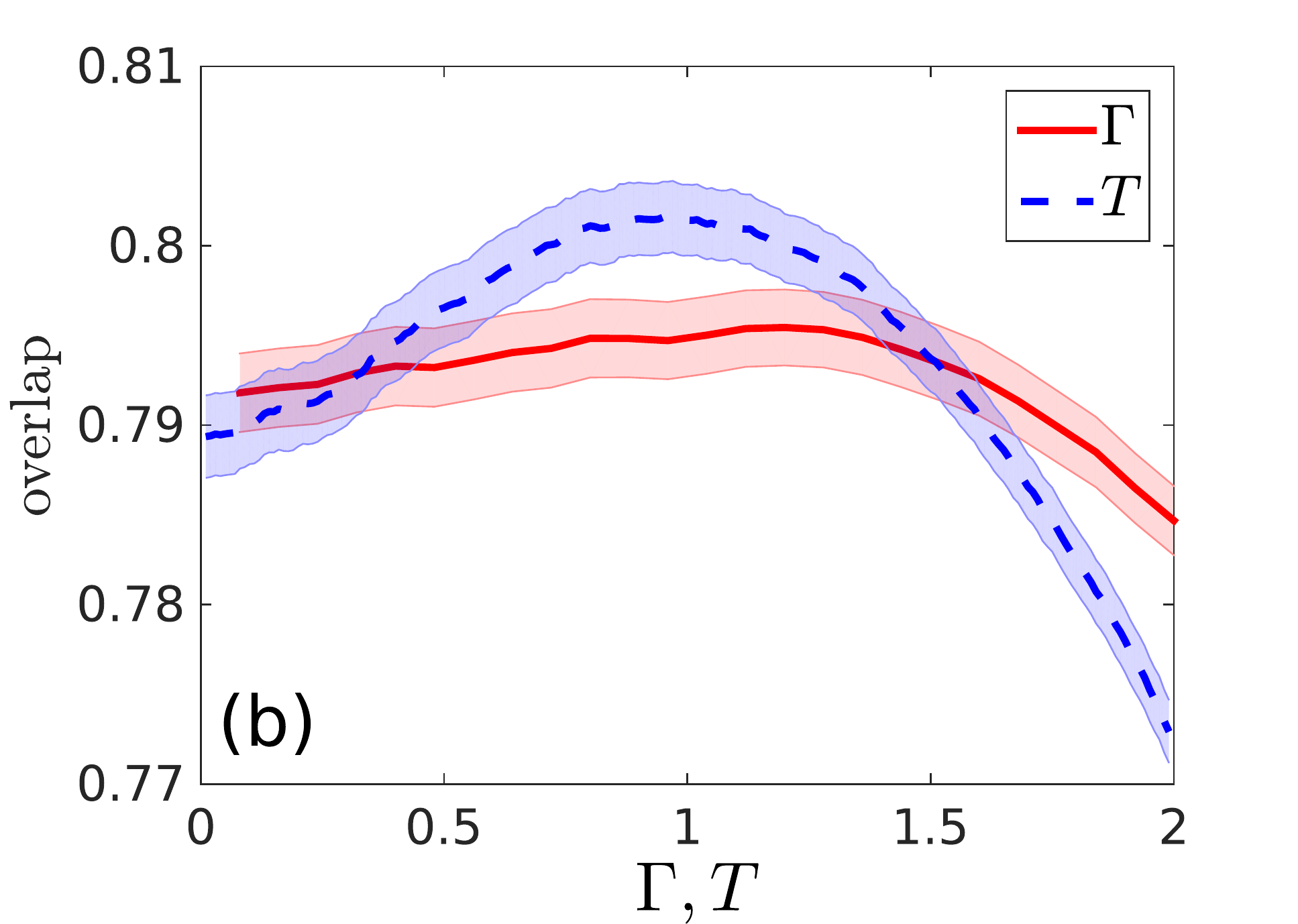}

	\caption{The overlap as a function of the transverse field $\Gamma$ (red solid line) and the temperature $T$ (blue dashed line, see \cite{Nishimura2016} for details) for (a) $\sigma=0.5, \gamma=1.0$ and (b) $\sigma=1.0, \gamma=0.4$. Standard deviations are indicated in the same color.}
	\label{fig:overlap}
\end{figure*}

The red solid line in Fig. \ref{fig:overlap} shows the results for two pairs of $\sigma$ and $\gamma$. The horizontal and vertical axes show the transverse field $\Gamma$ and the overlap, respectively. As seen in these figures, the overlap has a mild peak at a finite transverse field. This result indicates that it is better to keep the transverse field finite at the end of QA, in contrast to the conventional prescription, though the gain (the increase in $M$ compared to its value at $\Gamma=0$) is small. We also notice that the maximum value of the overlap is smaller for the present case than for the thermal case \cite{Nishimura2016}.

\begin{figure}
	\centering
	\includegraphics[width=\columnwidth]{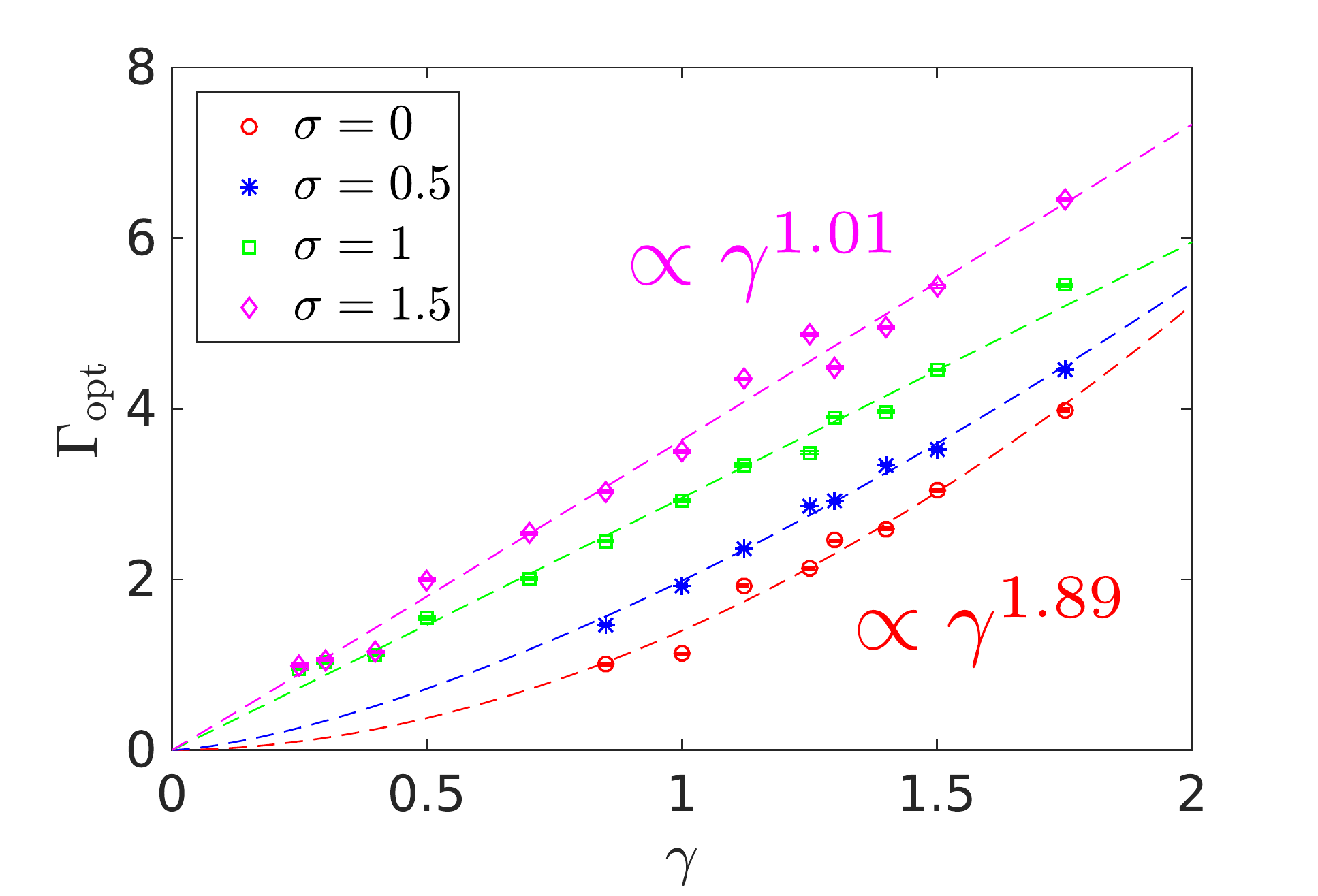}
	\caption{The optimal transverse magnetic field $\Gamma_{\mathrm{opt}}$ versus the strength of noise $\gamma$ for several values of $\sigma$. The dashed lines represent curve fittings.}
	\label{fig:maxtf}
\end{figure}

We next analyze the value of $\Gamma$ which maximizes the overlap $M(\Gamma)$ as a function of the noise strength $\gamma$. The result is in Fig. \ref{fig:maxtf}. The strength of noise and the corresponding optimal transverse magnetic field $\Gamma_{\mathrm{opt}}$ are indicated in the abscissa and the ordinate, respectively. Notice that if $\gamma = 0$, i.e., the situation where the interactions are not disturbed by noise, the optimal transverse magnetic field $\Gamma_{\mathrm{opt}}$ must be zero. Because the exact relation between $\gamma$ and $\Gamma_{\mathrm{opt}}$ is unknown, we apply a curve fitting to each plot, which is indicated as dotted lines in Fig. \ref{fig:maxtf}. These fittings show qualitatively that for small $\sigma$, $\Gamma_{\mathrm{opt}}$ behaves approximately quadratically whereas it is almost linear for large $\sigma$. 

\section{Mean-field analysis} 
\label{sec:mfa}

\begin{figure}
	\centering
	\includegraphics[width=\columnwidth]{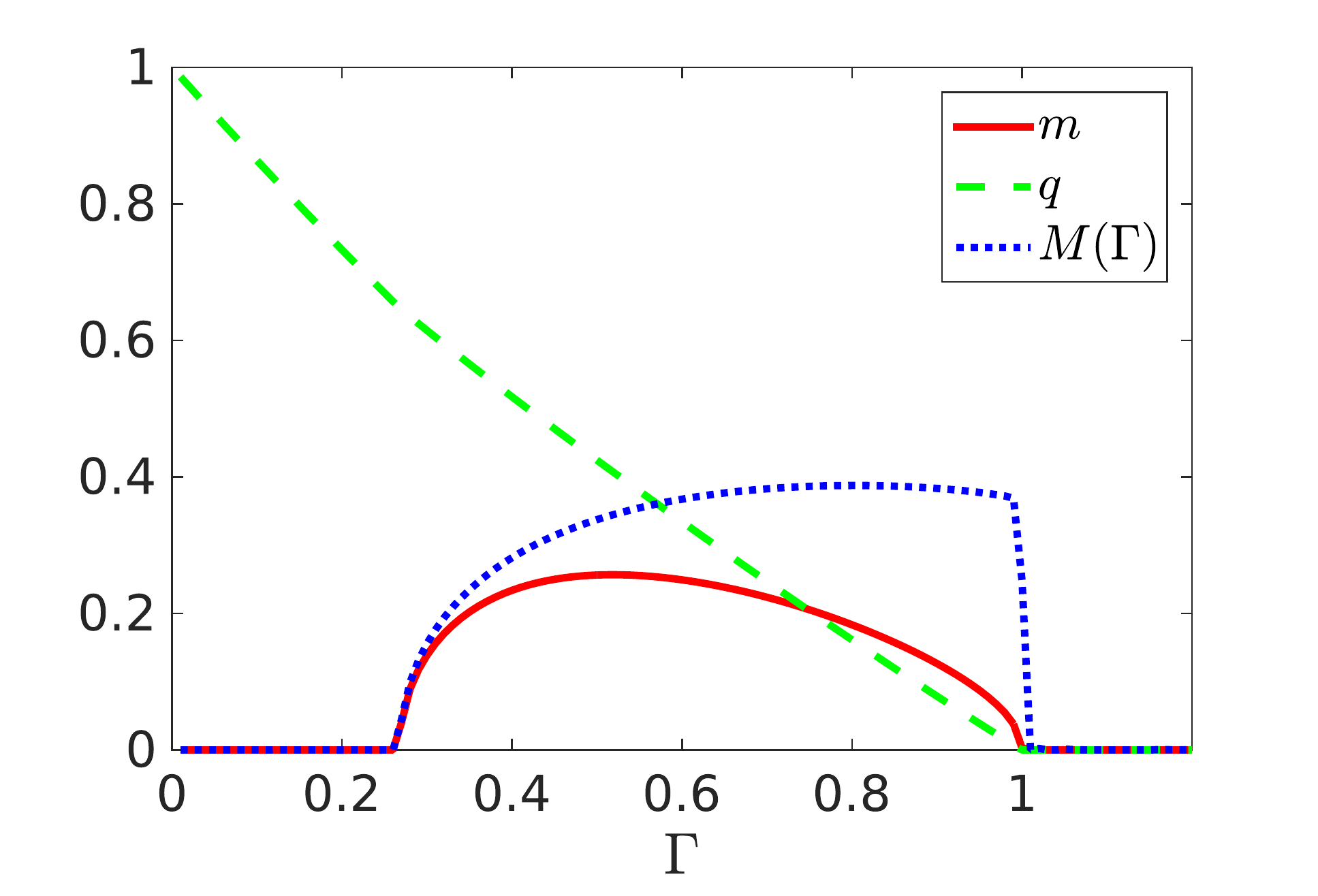}
	\caption{Shown dotted in blue is the overlap $M(\Gamma)$ as a function of $\Gamma$ for $\sigma=0.5$, $\gamma=0.75$, and $J_0 = 1$ at zero temperature. Also shown are the magnetization (full line in red) and the spin-glass order parameter (dashed line in green). Behavior at low $\Gamma$ may not be reliable due to replica symmetry breaking, which, however, would not change the existence of a peak in the overlap at an intermediate value of $\Gamma$.}
	\label{fig:overlapSK}
\end{figure}

To understand the behavior of the overlap qualitatively, we supplement the numerical analysis by the mean-field theory. The original Hamiltonian is the Sherrington-Kirkpatrick (SK) model \cite{skmodel,nsmrbook}, an Ising spin glass with full connectivity,
\begin{align}
	\label{eq:SK_S}
	H_{\mathrm{SK}}(\v{\hat{\Sig}}) = -\sum_{i<j}J_{ij}\hat{\Sig}_{i}^z\hat{\Sig}_{j}^z,
\end{align}
where the original interactions $J_{ij}$ are generated by the Gaussian distribution appropriately normalized for full connectivity,
\begin{align}
	P(J_{ij}) = \sqrt{\frac{N}{2\pi \sigma^2}}\exp\left\{-\frac{N}{2\sigma^2}\left(J_{ij}-\frac{J_0}{N}\right)^2\right\}
\end{align}
with mean $J_0$ and variance $\sigma^2$. The Hamiltonian with noise and transverse field is
\begin{align}
	\label{eq:SK_tau}
	\tilde{H}_{\mathrm{SK}}(\v{\hat{\tau}}, \Gamma) = -\sum_{i<j}\tilde{J}_{ij}\hat{\tau}_{i}^z\hat{\tau}_{j}^z - \Gamma \sum_{i}\hat{\tau}_{i}^x,
\end{align}
where $\hat{\tau}$ is the Pauli operator, and the interactions are degraded by noise $\xi_{ij}$,
\begin{align}
	\tilde{J_{ij}} = J_{ij}+\xi_{ij},
\end{align}
which follows the Gaussian distribution with zero mean and variance $\gamma^2$,
\begin{align}
	P(\xi_{ij}) = \sqrt{\frac{N}{2\pi \gamma^2}}\exp\left(-\frac{N}{2\gamma^2} \xi_{ij}^2\right).
\end{align}
The overlap equivalent to Eq. (\ref{eq:MGamma}) is defined as
\begin{align}
	M(\Gamma) = \lim_{\beta_0, \beta \to \infty} [\mathrm{sgn}\braket{\hat{\Sig}_{i}^z}_{\beta_0}\mathrm{sgn}\braket{\hat{\tau}_{i}^z}_{\beta, \Gamma}],
\end{align}
where the square brackets denote the configurational average over interactions $J_{ij}$ and noise $\xi_{ij}$,
\begin{align}
	\int \prod_{i<j}dJ_{ij}d\xi_{ij}P(J_{ij})P(\xi_{ij})(\cdots) = [\cdots],
\end{align}
and the angular brackets stand for the thermal average with respect to each Hamiltonian,
\begin{align}
	\braket{\hat{\Sig}_{i}^z}_{\beta_0} &= \frac{\mathrm{Tr}\hat{\Sig}_{i}^z \exp \left[-\beta_0 H_{\mathrm{SK}}(\v{\hat{\Sig}})\right]}{\mathrm{Tr}\exp \left[-\beta_0 H_{\mathrm{SK}}(\v{\hat{\Sig}})\right]} \\
	\braket{\hat{\tau}_{i}^z}_{\beta, \Gamma} &= \frac{\mathrm{Tr}\hat{\tau}_{i}^z \exp \left[-\beta \tilde{H}_{\mathrm{SK}}(\v{\hat{\tau}}, \Gamma)\right]}{\mathrm{Tr}\exp \left[-\beta \tilde{H}_{\mathrm{SK}}(\v{\hat{\tau}}, \Gamma)\right]}. 
\end{align}
The task is to find the optimal transverse field $\Gamma$ to maximize the overlap. As detailed in the Appendix, the free energy per spin is calculated by the Suzuki-Trotter decomposition under the ansatz of replica symmetry and the static approximation. Before taking the limit of large $\beta_0$ and $\beta$, this free energy is 
\begin{align}
	\label{eq:feps}
	\begin{split}
		-[f] &= \frac{\sigma^2 \beta_0^2}{4}q_0^2 + \frac{\sigma^2 \beta_0^2}{4} + \frac{(\sigma^2+\gamma^2)\beta^2}{4}(q^2-r^2)  \\
		&-\frac{J_0 \beta_0}{2}m_0^2-\frac{J_0 \beta}{2}m^2-\frac{\sigma^2 \beta_0^2 q_0^2}{2} \\
		&+ \int Dz_0 \ln(2\cosh{\Phi_0})  \\
		&+ \int Dz \ln\left(\int Dy \ 2\cosh{\Psi}\right),
	\end{split}
\end{align}
where
\begin{align}
	\Phi_0(z_0) &= \sqrt{\sigma^2 \beta_0^2 q_0}z_0 + J_0 \beta_0 m_0, \\
	\Phi(z,y) &= \sqrt{(\sigma^2+\gamma^2) \beta^2 q}z + J_0 \beta m \nonumber \\&+ \sqrt{(\sigma^2+\gamma^2)\beta^2(r-q)}y, \\
	\Psi &= \sqrt{\beta^2\Gamma^2+\Phi^2}.
\end{align}
The order parameters $m_0, q_0, m, q, $ and $r$ satisfy the self-consistent equations,
\begin{align}
	m_0 &= \int Dz_0 \tanh{\Phi_0} = [\braket{\hat{\Sig}_i^z}_{\beta_0}], \\
		q_0 &= \int Dz_0 \tanh^2{\Phi_0} = [\braket{\hat{\Sig}_i^z}_{\beta_0}^2], \\
		m &= \int Dz \frac{\int Dy \left(\frac{\Phi}{\Psi}\right)\sinh{\Psi}}{\int Dy \cosh \Psi} = [\braket{\hat{\tau}_{i}^z}_{\beta, \Gamma}], \\
		q &= \int Dz \left\{\frac{\int Dy \left(\frac{\Phi}{\Psi}\right)\sinh{\Psi}}{\int Dy \cosh \Psi}\right\}^2 = [\braket{\hat{\tau}_{i}^z}_{\beta, \Gamma}^2], \\
		r &= \int Dz \frac{\int Dy \left\{\left(\frac{\Phi}{\Psi}\right)^2\cosh{\Psi}+\left(\frac{\beta^2 \Gamma^2}{\Psi^3}\right) \sinh{\Psi}\right\}}{\int Dy \cosh \Psi}.
\end{align}
As was the case in the classical problem \cite{Nishimura2016}, the two systems, Eqs. (\ref{eq:SK_S}) and (\ref{eq:SK_tau}), decouple under the ansatz we used. We hence obtain 
\begin{align}
	\label{eq:MGammaSK}
	M(\Gamma) &= \lim_{\beta_0, \beta \to \infty} [\mathrm{sgn}\braket{\hat{\Sig}_{i}^z}_{\beta_0}\mathrm{sgn}\braket{\hat{\tau}_{i}^z}_{\beta, \Gamma}] \nonumber \\
	&= \lim_{\beta_0, \beta \to \infty} [\mathrm{sgn}\braket{\hat{\Sig}_{i}^z}_{\beta_0}]_{\sigma^2}[\mathrm{sgn}\braket{\hat{\tau}_{i}^z}_{\beta, \Gamma}]_{\sigma^2+\gamma^2},
\end{align}
where the subscripts $\sigma^2$ and $\sigma^2+\gamma^2$ denote the variance of randomness. Since the first factor of Eq. (\ref{eq:MGammaSK}), $[\mathrm{sgn}\braket{\hat{\Sig}_{i}^z}_{\beta_0}]_{\sigma^2}$, is independent of $\Gamma$, we focus our attention only on the second factor $[\mathrm{sgn}\braket{\hat{\tau}_{i}^z}_{\beta, \Gamma}]_{\sigma^2+\gamma^2}$. After taking the limit of large $\beta$, we plot the overlap as a function of $\Gamma$ in Fig. \ref{fig:overlapSK}.
As seen in this figure, the overlap has a  maximum at a finite $\Gamma$ \footnote{The behaviors of the overlap and the order parameters drawn in this figure are not reliable at low temperatures due to the replica symmetry breaking.  Nevertheless, the existence of a peak in the overlap is likely to be established since the peak position is at a relatively high temperature.}.

\begin{figure*}
	\centering
	
	\includegraphics[width=\columnwidth]{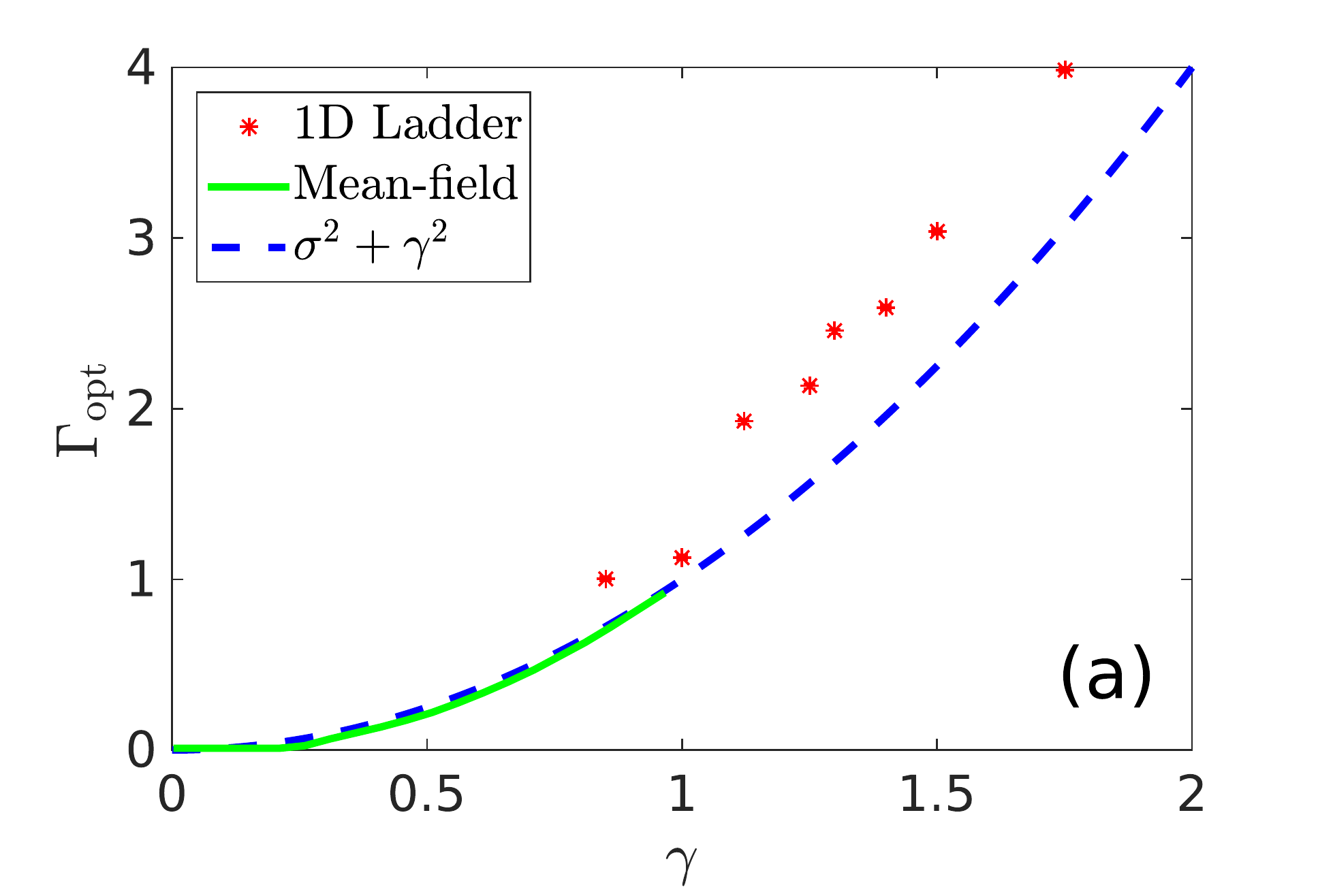}
	\includegraphics[width=\columnwidth]{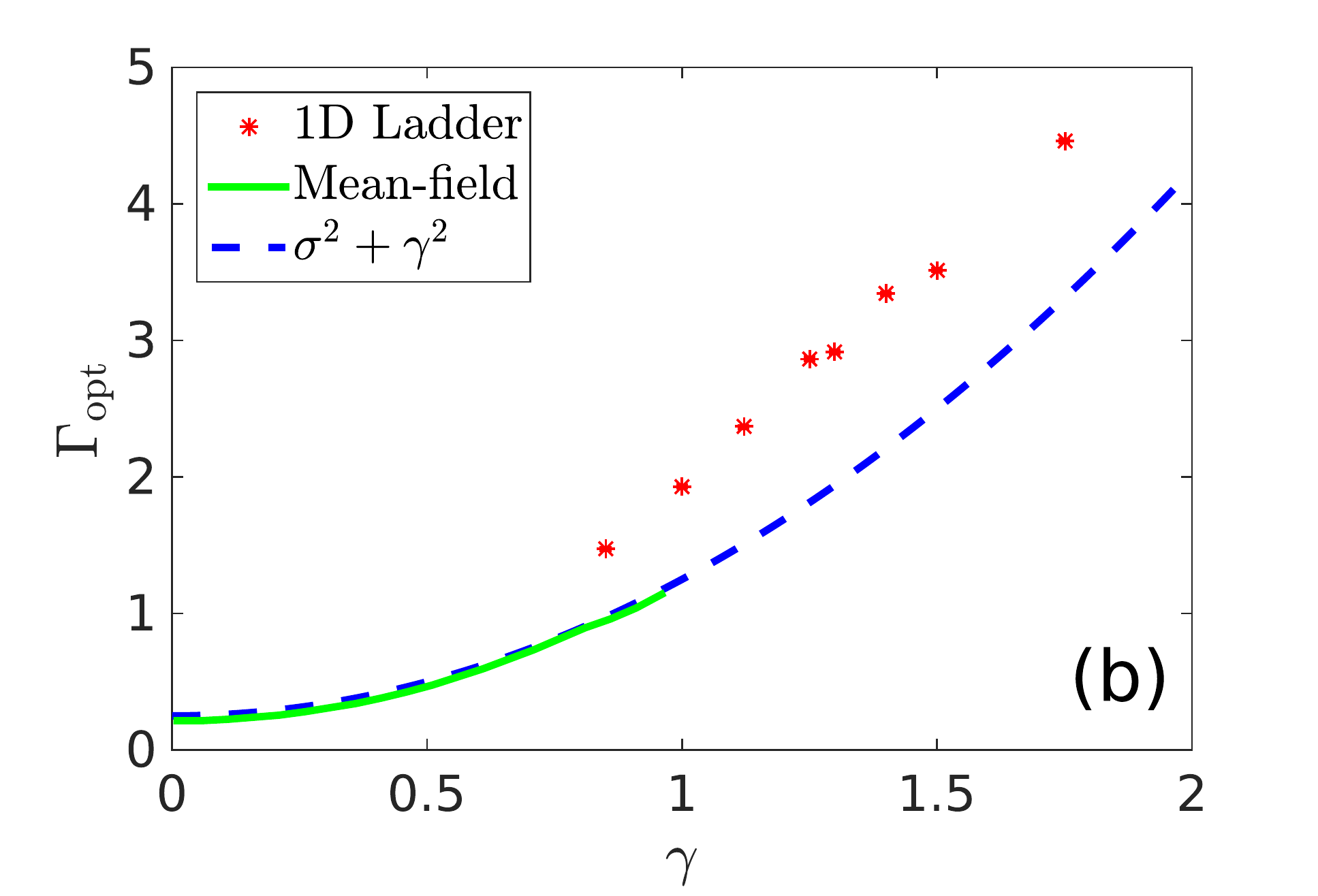}

	\caption{The green solid line is the optimal transverse field $\Gamma_{\mathrm{opt}}$ as a function of $\gamma$ for the SK model obtained by solving the self-consistent equations numerically. The blue dashed line denotes $\sigma^2+\gamma^2$. The red dots are for $\Gamma_{\mathrm{opt}}$ on triangular 1D ladder (cf. Fig. \ref{fig:maxtf}). Panels (a) and (b) are for $\sigma = 0$ and $\sigma = 0.5$, respectively.}
	\label{fig:Gopt}
\end{figure*}

The green solid line in Fig. \ref{fig:Gopt} depicts the optimal transverse field $\Gamma_{\mathrm{opt}}$ as a function of noise $\gamma$. We find that  $\Gamma_{\mathrm{opt}}$ closely follows the curve $\sigma^2 + \gamma^2$ (the blue dashed line in Fig. \ref{fig:Gopt}). For this reason, we may conjecture that the analytical form of $\Gamma_{\mathrm{opt}}$ is close to $\sigma^2+\gamma^2$, which is the same expression as the optimal temperature obtained in our previous study of the classical case \cite{Nishimura2016}. The results for the quasi-one-dimensional triangular ladder are also plotted in Fig. \ref{fig:Gopt} in red dots.  The mean-field curve behaves qualitatively similarly but more work is needed to achieve quantitative understanding.

\section{Conclusion} 
\label{sec:conclusion}

We have studied the problem of quantum-mechanical inference of the original ground state of the Ising spin glass from the Hamiltonian whose interactions are degraded by noise. We first applied a numerical method, DMRG, to a quasi-one-dimensional system on the triangular ladder in a transverse field. It has been found that there exists a finite strength of the transverse field that maximizes the overlap. We next used the mean-field model to investigate the behavior of the overlap analytically. Under the replica symmetric and static approximations, the optimal transverse field was calculated from the self-consistent equations. It has been shown that a finite value of the transverse field gives a larger overlap than the zero-field case, as in the quasi-one-dimensional problem.  We may therefore reasonably guess that similar behavior would be expected for more generic cases.

The absolute value of the peak of the overlap is smaller than in the classical case as seen in Fig. \ref{fig:overlap} though the difference is not very significant, of the order of $1\%$. The present method has an advantage over the classical thermal approach that the strength of the transverse field is a controllable parameter on the hardware, whereas the temperature is kept fixed.

A possible drawback is that the value of the optimal transverse field is hard to predict beforehand, though there exist some hints as indicated in Fig. \ref{fig:Gopt}.  We, nevertheless, observe in Figs. \ref{fig:overlap} and \ref{fig:overlapSK} that the dependence of the overlap on $\Gamma$ is relatively mild around the peak, which implies that the result would not change significantly if we stop the process of QA near, but not exactly equal to, the optimal value of $\Gamma$. At least, we would expect a better result at a small $\Gamma$ than at $\Gamma =0$.

It is not easy to explain intuitively why quantum fluctuations are useful to improve the performance of ground-state inference out of a noisy Hamiltonian.  A finite value of the transverse field would change the energy landscape, which may partly compensate for the noise in interactions.  More work is needed to better understand the mechanism behind the non-monotonic behavior of the overlap as a function of the transverse field.

\begin{acknowledgments}
This work was funded partly by the ImPACT Program of Council for Science, Technology and Innovation, Cabinet Office, Government of Japan, and by the JPSJ KAKENHI Grant No. 26287086.
\end{acknowledgments}

\appendix
\section{Mean-field free energy}
In this Appendix, we show the derivation of the free energy per spin Eq. (\ref{eq:feps}) for the SK model.

First we apply the Suzuki-Trotter decomposition \cite{Trotter1959, Suzuki1976} to the partition function so that we can describe the partition function with effective classical Hamiltonians $H_{\mathrm{eff}}(\v{\Sig})$ and $H_{\mathrm{eff}}(\v{\tau})$,
\begin{align}
	Z(\beta_0, \beta, \Gamma) &= \mathrm{Tr} \exp{\left(-H_{\mathrm{eff}}(\v{\Sig})-\tilde{H}_{\mathrm{eff}}(\v{\tau})\right)}, \\
	H_{\mathrm{eff}}(\v{\Sig}) &= -\sum_{i<j}\beta_0 J_{ij}\Sig_i \Sig_j,\\
	\tilde{H}_{\mathrm{eff}}(\v{\tau}) &= \sum_{i<j}\sum_{t=1}^{P}\frac{\beta}{P}\tilde{J}_{ij}\tau_i(t)\tau_j(t) \nonumber \\
	&-\sum_{i}\sum_{t=1}^{P}B\tau_i(t)\tau_i(t+1),
\end{align}
where $P$ shows the trotter number and $B = (1/2)\log{\coth{(\beta \Gamma/P)}}$. Following the standard prescription of replica method \cite{nsmrbook}, the $n$th power of the partition function is written as
\begin{widetext}
\begin{align}
	[Z^n] &= \int \prod_{i<j}dJ_{ij}d{\xi_{ij}}P(J_{ij})P(\xi_{ij}) \left\{\mathrm{Tr}\exp{\left(-H_{\mathrm{eff}}(\v{\Sig})-\tilde{H}_{\mathrm{eff}}(\v{\tau})\right)}\right\}^{n} \nonumber \\
	&= \int \prod_{i<j}dJ_{ij}d{\xi_{ij}}P(J_{ij})P(\xi_{ij}) \mathrm{Tr}\exp\left(\sum_{\alpha}\sum_{i<j}\beta_0 J_{ij}\Sig_i^{\alpha}\Sig_j^{\alpha}+\sum_{a,i,t}B\tau_i^{\alpha}(t)\tau_i^{\alpha}(t+1)\right. \nonumber \\
	&+\left.\sum_{\alpha}\sum_{i<j}\sum_{t}\frac{\beta}{P}(J_{ij}+\xi_{ij})\tau_i^{\alpha}(t)\tau_j^{\alpha}(t)\right).
\end{align}
Using the replica method, we obtain the free energy density as
\begin{align}
	-[f] &= \lim_{n \to 0} \left(-\frac{\sigma^2 \beta_0^2}{2n}\sum_{\alpha < \beta}{q_{\alpha \beta}^{0}}^2+\frac{\sigma^2 \beta_0^2}{4}-\frac{(\sigma^2+\gamma^2)\beta^2}{2P^2n}\sum_{\alpha<\beta,tt'}q_{\alpha \beta}(t,t')^2-\frac{(\sigma^2+\gamma^2)\beta^2}{4P^2n}\sum_{\alpha,tt'}q_{\alpha \alpha}(t,t')^2\right. \nonumber \\
	&-\left.\frac{\sigma^2 \beta_0 \beta}{2Pn}\sum_{\alpha\beta,t}u_{\alpha \beta}(t)^2-\frac{J_0 \beta_0}{2n}\sum_{\alpha}{m_{\alpha}^{0}}^2-\frac{J_0 \beta}{2Pn}\sum_{\alpha,t}m_{\alpha}(t)^2+\frac{1}{n}\log{\mathrm{Tr}e^{L}}\right),
\end{align}
where
\begin{align}
	L &\equiv \sigma^2 \beta_0^2 \sum_{\alpha<\beta}q_{\alpha \beta}^{0}\Sig^{\alpha}\Sig^{\beta}+\frac{(\sigma^2+\gamma^2)\beta^2}{P^2}\sum_{\alpha<\beta,tt'}q_{\alpha \beta}(t,t')\tau^{\alpha}(t)\tau^{\beta}(t')+\frac{(\sigma^2+\gamma^2)\beta^2}{2P^2}\sum_{\alpha,tt'}q_{\alpha \alpha}(t,t')\tau^{\alpha}(t)\tau^{\alpha}(t') \nonumber \\
	&+\frac{\sigma^2\beta_0 \beta}{P}\sum_{\alpha \beta,t}u_{\alpha \beta}(t)\Sig^{\alpha}\tau^{\beta}(t)+J_0 \beta_0 \sum_{\alpha}m_{\alpha}^{0}\Sig^{\alpha}+\frac{J_0 \beta}{P}\sum_{\alpha,t}m_{\alpha}(t)\tau^{\alpha}(t)+\sum_{\alpha,t}B\tau^{\alpha}(t)\tau^{\alpha}(t+1).
\end{align}
The order parameters are determined by the saddle-point conditions,
\begin{align}
	\begin{cases}
		q_{\alpha \beta}^{0}  &= \braket{\Sig^{\alpha}\Sig^{\beta}}_{L} = \left[\braket{\Sig_{i}^{\alpha}}_{\beta_0}\braket{\Sig_{i}^{\beta}}_{\beta_0}\right], \\
		q_{\alpha \beta}(t,t')&= \braket{\tau^{\alpha}(t)\tau^{\beta}(t')}_{L} = \left[\braket{\tau_i^{\alpha}(t)}_{\beta,\Gamma}\braket{\tau_i^{\beta}(t)}_{\beta,\Gamma}\right], \\
		m_{\alpha}^{0}  &= \braket{\Sig^{\alpha}}_{L} = \left[\braket{\Sig_{i}^{\alpha}}_{\beta_0}\right], \\
		m_{\alpha}(t)&= \braket{\tau^{\alpha}(t)}_{L} = \left[\braket{\tau_i^{\alpha}(t)}_{\beta,\Gamma}\right], \\
		u_{\alpha \beta}(t)&= \braket{\Sig^{\alpha}\tau^{\beta}(t')}_{L} = \left[\braket{\Sig_{i}^{\alpha}}_{\beta_0}\braket{\tau_i^{\beta}(t)}_{\beta,\Gamma}\right], \\
	\end{cases}
\end{align}
where
\begin{align}
	\braket{\cdots}_{L} = \frac{\mathrm{Tr}(\cdots)e^{L}}{\mathrm{Tr}e^{L}}.
\end{align}
If we use the replica symmetric ansatz and static approximation, we have
\begin{align}
	q_{\alpha \beta}^{0}=
	\begin{cases}
		q_0 & [\alpha \neq \beta] \\
		0 & [\alpha = \beta]
	\end{cases},\ 
	q_{\alpha \beta}(t,t')=
	\begin{cases}
		q & [\alpha \neq \beta] \\
		r & [\alpha = \beta]
	\end{cases},\ 
	m_{\alpha}^{0} = m_0,\  m_{\alpha}(t) = m,\  u_{\alpha \beta}(t) = u.
\end{align}
The free energy per spin is reduced to
\begin{align}
	-[f] = \frac{\sigma^2 \beta_0^2}{4}{q_0}^{2}+\frac{\sigma^2 \beta_0^2}{4}+\frac{(\sigma^2+\gamma^2)\beta^2}{4}(q^2-r^2)-\frac{J_0 \beta_0}{2}{m_0}^2-\frac{J_0 \beta}{2}m^2+\lim_{n \to 0}\log{\mathrm{Tr}e^{L_0}},
\end{align}
where $L_0$ represents $L$ under the replica symmetric ansatz and the static approximation,
\begin{align}
	\label{eq:logtrexpL0}
	\log{\mathrm{Tr}e^{L_0}} &= \log\mathrm{Tr}\exp\left(\sigma^2 \beta_0^2 q_0\sum_{\alpha < \beta}\Sig^{\alpha} \Sig^{\beta}+\frac{(\sigma^2+\gamma^2)\beta^2 q}{P^2}\sum_{\alpha<\beta,tt'}\tau^{\alpha}(t)\tau^{\beta}(t')+\frac{(\sigma^2+\gamma^2)\beta^2 r}{2 P^2}\sum_{\alpha,tt'}\tau^{\alpha}(t)\tau^{\alpha}(t')\right. \nonumber \\
	&\left.+\frac{\sigma^2 \beta_0 \beta u}{P}\sum_{\alpha\beta,t}\Sig^{\alpha}\tau^{\beta}(t)+J_0 \beta_0 m_0\sum_{\alpha} \Sig^{\alpha}+\frac{J_0 \beta m}{P}\sum_{\alpha,t}\tau^{\alpha}(t)+\sum_{\alpha,t}B\tau^{\alpha}(t)\tau^{\alpha}(t+1)\right).
\end{align}
By using the formula
	\footnote{Note that $Dm \equiv (1/\sqrt{2\pi})\exp{(-m^2/2)} dm$ denotes the Gaussian measure.},
\begin{align}
	\exp{\left(\frac{ax^2}{2}\right)} = \int Dm e^{\sqrt{a}mx},
\end{align}
Eq. (\ref{eq:logtrexpL0}) can be simplified as
\begin{align}
	\log{\mathrm{Tr}e^{L_0}} &= \log \int Dz_0 Dz Dw \left(2\cosh{\Phi_0} \cdot \int Dy 2\cosh{\sqrt{\beta^2\Gamma^2+\Phi^2}}\right)^n-\frac{\sigma^2 \beta_0^2 q_0 n}{2} \nonumber \\
	&=n\left\{\int Dz_0 Dz Dw \log\left(2\cosh{\Phi_0} \cdot \int Dy 2\cosh{\sqrt{\beta^2\Gamma^2+\Phi^2}}\right)-\frac{\sigma^2 \beta_0^2 q_0}{2}\right\}+\mathcal{O}(n^2),
\end{align}
where
\begin{align}
	\Phi_0 &= \sqrt{a_0}z_0 + b_0 + \sqrt{c}w, \\
	\Phi &= \sqrt{a}z + b + \sqrt{c}w + \sqrt{d}y,
\end{align}
and
\begin{align}
	\begin{cases}
		a_0 &= \sigma^2 \beta_0^2 q_0 - \sigma^2 \beta_0 \beta u \\
		a &= (\sigma^2+\gamma^2) \beta^2 q - \sigma^2 \beta_0 \beta u \\
		b_0 &= J_0 \beta_0 m_0 \\
		b &= J_0 \beta m \\
		c &= \sigma^2 \beta_0 \beta u \\
		d &= (\sigma^2+\gamma^2)\beta^2 (r-q) \\
	\end{cases}.
\end{align}
A straightforward calculation shows that the free energy does not depend on $u$, similarly to the classical case \cite{Nishimura2016},
\begin{align}
	-\pd{[f]}{u} = 0.
\end{align}
Therefore, the $w$-dependence of $\Phi_0$ and $\Phi$ disappears and we finally obtain Eq. (\ref{eq:feps}).
Consequently, the explicit form of the overlap is derived as
\begin{align}
	M(\Gamma) = \lim_{\beta_0, \beta \to \infty}\int Dz_0 \mathrm{sgn \Phi_0}\int Dz \mathrm{sgn}\left(\frac{\int Dy \sinh \sqrt{\beta^2 \Gamma^2+\Phi^2}\cdot \frac{\Phi}{\beta^2 \Gamma^2 + \Phi^2}}{\int Dy \cosh \sqrt{\beta^2 \Gamma^2 + \Phi^2}}\right).
\end{align}
\end{widetext}

\bibliographystyle{apsrev4-1}
\bibliography{./reference.bib}

\end{document}